\documentclass[12pt]{article}
\usepackage{amssymb,amsmath,slashed,latexsym}
\newcommand{\ft}[2]{{\textstyle\frac{#1}{#2}}}
\def\Re{\mathop{\rm Re}\nolimits}

\def\rme{{\rm e}}
\def\rmi{{\rm i}}

\newcommand{\hc}{{\rm h.c.}}
\newsavebox{\uuunit}
\sbox{\uuunit}
    {\setlength{\unitlength}{0.825em}
     \begin{picture}(0.6,0.7)
        \thinlines
        \put(0,0){\line(1,0){0.5}}
        \put(0.15,0){\line(0,1){0.7}}
        \put(0.35,0){\line(0,1){0.8}}
       \multiput(0.3,0.8)(-0.04,-0.02){12}{\rule{0.5pt}{0.5pt}}
     \end {picture}}
\newcommand {\unity}{\mathord{\!\usebox{\uuunit}}}

\csname @addtoreset\endcsname{equation}{section}
\newcommand{\SU}{\mathop{\rm SU}}
\newcommand{\SO}{\mathop{\rm SO}}
\newcommand{\U}{\mathop{\rm {}U}}

   \usepackage[nosort]{cite}
   \pdfoutput=1
  \usepackage[pdftex]{hyperref}
\begin{document}

\begin{titlepage}
\phantom{.}
\vspace{.5cm}
\begin{center}
\baselineskip=16pt
{\LARGE    Matter Couplings in Supergravity  \\ \vskip 0.2cm 
\large The first 10 years}\\
\vfill
{\large Antoine Van Proeyen  
  } \\
\vfill
{\small KU Leuven, Institute for Theoretical Physics,\\[2mm]
       and
      \\[2mm] Leuven Gravity Institute, KU Leuven, \\[2mm]
      Celestijnenlaan 200D, box 2415, B-3001 Leuven, Belgium
}
\end{center}
\vfill
\begin{center}
{\bf Abstract}
\end{center}
{\small Following the initial construction of pure supergravity in 1976, various methods were developed to couple supergravity with supersymmetric matter. This contribution to “Half a century of supergravity" provides a personal perspective on the key steps, techniques and results developed in the first decade. These developments form the foundation for numerous applications in phenomenology, cosmology and string theories, while also revealing intriguing mathematical structures. }\vspace{2mm} \vfill 

\noindent {\em Invited contribution to “Half a century of Supergravity”\\
eds. A. Ceresole and G. Dall’Agata (Cambridge Univ. Press, to appear)}

\end{titlepage}
\addtocounter{page}{1}
 \tableofcontents{}
\newpage
\section{Introduction}

Early 1976, the first supergravity action was obtained in  \cite{Freedman:1976xh}, and shortly thereafter in  \cite{Deser:1976eh}. 
 An immediate goal was to couple all known supersymmetric theories to supergravity. Any field theory with rigid Poincar\'{e} symmetry can be upgraded to a theory with Einstein invariance, i.e. invariant under local general coordinate transformations, by coupling it to the graviton field. This transition to a locally invariant theory follows a relatively direct path. Similarly, it was expected that globally supersymmetric matter actions could be coupled to the spin-(2,3/2) gauge fields to achieve locally supersymmetric actions. However, this procedure is less straightforward than for general relativity. The transformation laws of the supergravity multiplet and the supersymmetric matter multiplets in general mix. This makes the process less straightforward but also much more interesting, as it produces a richer and more intricate structure. The resulting theories unveiled a wealth of deep mathematical structures that continue to inspire research today. This review will outline the various steps that our colleagues have followed, and the structures they revealed in the last half century.

This contribution to 'Half a century of Supergravity' is a historic account of the early developments in supergravity-matter couplings. I was not yet involved in the early days of supergravity. My first contact was a seminar by Peter van Nieuwenhuizen in Leuven, 
followed by the summer school in Carg\`{e}se in July 1978. But I have reconstructed these early developments from papers and insights shared by my friends, whose work I came to appreciate soon after. Writing this review reinforced my admiration for how much was accomplished in the short time following the discovery of supergravity. Since multiple research directions were followed from the start, I could not write everything in the historical order. Where possible, I will follow a historical thread and add how I personally learned to appreciate different contributions. As the title of this review indicates, I concentrate on the matter couplings. This research was not independent from other major advances, such as extended supergravities, higher dimensions, quantum supergravity, supersymmetric black holes and their attractor mechanism, \ldots, that I will not discuss here but that you will find in other parts of this  book.

Of course, supergravity-matter couplings are based on progress in understanding rigid supersymmetry, which I will also not review here. In the following section, I begin with the elementary  Noether coupling method, which was predominantly used in the first year, 1976. Significant progress was made once auxiliary fields were found, allowing supersymmetry transformations to be formulated independently of the specific action under consideration (see Sec.~\ref{ss:auxfields}). While many later developments occurred simultaneously, I will discuss them in separate sections.

Sec.~\ref{ss:tensorcalculus} reviews the method of combining off-shell multiplets, called `tensor calculus', which was a major breakthrough in 1978. That summer, the Carg\`{e}se school (July 1978) provided a first main encounter of supergravity practitioners. The main lectures from that school will also be discussed in Sec.~\ref{ss:tensorcalculus}. Around the same time superspace methods for supergravity were being developed, a topic covered in Sec.~\ref{ss:superspace}, which also highlights the crucial role that the 1979 Stony Brook supergravity meeting played in the early development of supergravity.

The central discussion of general matter couplings to ${\cal N}=1$ supergravity is found in Sec.~\ref{ss:gencoupl_susybreaking}. Supersymmetry breaking had already been a topic of interest long before (1973–77), which I will first review, before turning to the formulation of general ${\cal N}=1$ matter couplings. This was where my own contributions began. The key results were obtained in 1978-82. These developments also fueled the attempts for realistic models, which contain hidden and observable sectors. A particular type of such models, no-scale supergravities, became very popular in 1983-85 and appear at the end of that section.

The results became more transparent with the help of the superconformal tensor calculus. Its development had already begun in 1976 and in Sec.~\ref{ss:superconform} the main steps of superconformal calculus are reviewed.
Simultaneously, also the group manifold approach (see Sec.~\ref{groupmanifold}) was being developed, starting in 1978 and continuing to evolve over the subsequent years.

You might find it surprising in a review on matter couplings that geometries like K\"{a}hler manifolds do not appear earlier. Nevertheless, that is how the field unfolded. The developments on K\"{a}hler manifolds are discussed in Sec.~\ref{ss:geometry}. The significance of K\"{a}hler geometry in supergravity emerged in works such as Bruno Zumino’s papers in 1979, discussions at the 1980 Erice meeting (which I will cover extensively), and the influential Bagger-Witten papers of 1982.

Sec.~\ref{ss:N2} shifts focus to ${\cal N}=2$ supersymmetry matter couplings. To properly introduce this topic, we must backtrack slightly, as the first models, auxiliary fields, and tensor calculus for ${\cal N}=2$ were explored between 1977 and 1982. The general theory was mostly found in 1983-84. These studies also led to the development of harmonic and projective superspace. The discussion on partial supersymmetry breaking ${\cal N}=2\to{\cal N}=1$ also started in 1984-85.

The general couplings of ${\cal N}=2$  supergravity gave rise to the concept of special geometry, discussed in Sec.~\ref{ss:quatSpecGeom}. The first ideas were developed in 1983-85, and it became hot in 1985 due to its relevance for Calabi-Yau compactifications of $D=10$ supergravities. By this point, we are approaching the final year covered in this review, 1985, which marked the beginning of extensive research into special quaternionic and K\"{a}hler (and real for $D=5$) geometries.

Sec.~\ref{ss:N4D56} addresses ${\cal N}=4$ supergravity in $D=4$, as well as matter couplings in $D= 5$ and $D=6$ dimensions. ${\cal N}=4$ supergravity had been studied in 1977, but its matter couplings only began to be explored in 1980.
The (minimally supersymmetric) $D=5$ and $D=6$ matter couplings are very similar and related to those in ${\cal N}=2,\,D=4$ supergravities. They were investigated in the final years covered in this review, and in many years later. They contributed to a broader picture of `special geometries'.

Finally, Sec.~\ref{ss:conclusions} reflects on later developments and their impact on phenomenology, cosmology and string theories, and finishes with some final thoughts.

\section{Noether method}
\label{ss:Noether}

In the months following the discovery of pure supergravity, significant results were quickly obtained by elementary methods, particularly by coupling the Noether currents of supersymmetry to the gravitino.
As early as August and September of 1976, the first example of matter coupling was achieved in  \cite{Ferrara:1976um,Freedman:1976ej,deWit:1976sk,Freedman:1976uk}:
the coupling of a vector multiplet - consisting of a spin-1 field and its spin-1/2 partner - to supergravity: a supersymmetric Maxwell-Einstein theory.

Just weeks later, a coupling of a spin (1/2,0,0) scalar multiplet to the (2,3/2) gauge multiplet has been obtained \cite{Ferrara:1976ni,Ferrara:1976kg}, followed shortly by studies of other interactions \cite{Cremmer:1977za,Das:1977mg,Das:1977pu}.

It has been a remarkable year. The invention of supergravity sparked an explosion of research in various directions. That explosion set a powerful movement in motion, one that could no longer be stopped. By the end of the year, already a lot was known about matter couplings.

The general approach used to construct such theories is known as the Noether method. One starts from putting a rigid supersymmetric theory in curved space, i.e. couple it to the graviton. Then couples the gravitino to the Noether current of supersymmetry, and adds the pure supergravity action.  Finally, one adds (laboriously and painfully) additional terms, order by order in the gravitational coupling constant $\kappa $, to the action and transformation laws to achieve full local supersymmetry invariance.

Also the first ${\cal N}$-extended supergravities were found in such a way:  ${\cal N}=2$  \cite{Ferrara:1976fu,Freedman:1976aw,Ferrara:1976iq} and  ${\cal N}=4$ \cite{Das:1977uy,Cremmer:1977tc,Cremmer:1977tt}.


 It was soon recognized that more effective methods were needed. These not only facilitated the construction of all interactions but also provided deeper insight into the structure of supergravity with matter. These improved methods will be discussed below, but let us first turn to the early days.

\section{Auxiliary fields}
\label{ss:auxfields}

With the Noether method, the transformations of the fields had always to be adapted, step by step, to the situation at hand. Determining the appropriate transformations was an integral part of constructing the theory, order by order in $\kappa$. The commutator of the resulting transformations yielded in general only to other transformations modulo the field equations of the constructed theory.

It became clear that the construction of matter couplings could be significantly simplified if `auxiliary fields' would be known: additional fields that, while not carrying physical degrees of freedom, ensure that the supersymmetry algebra closes without relying on equations of motion.
Such auxiliary fields were already well known in rigid supersymmetry.
When equations of motion are required to satisfy the supersymmetry algebra it is called an `open algebra'. Conversely, when the algebra of symmetries is satisfied without the need for equations of motion, it is called a `closed algebra'.

An early result of Peter Breitenlohner introduced an auxiliary field formulation with many components \cite{Breitenlohner:1976nv,Breitenlohner:1977jn}: 20 bosonic  and 20 fermionic. 
However, this set was not easy to work with. A `minimal set of auxiliary fields' was discovered nearly simultaneously at CERN, in London, and in Moscow \cite{Ferrara:1978em,Stelle:1978ye,Fradkin:1978jq}. 
These auxiliary fields for pure supergravity consist of two real scalars and a vector.
When added to the graviton and gravitino, they form a field set with 12+12 components (`supergravity multiplet') on which local supersymmetry transformations are defined. This has been called the `supergravity multiplet with old minimal set of auxiliary fields'. Three years later, another formulation with only 12+12 components, known as the `new minimal sets of auxiliary fields', was found \cite{Sohnius:1981tp}.  
In this case, a gauge vector and a gauge antisymmetric tensor are added to the physical fields.
Later, other sets were explored, such as the 16+16 multiplet in \cite{Girardi:1984vq}. 
However, the `old' and `new' minimal sets remained the most useful for discussing matter couplings.

The multiplets of rigid supersymmetry were then constructed in a locally supersymmetric version by incorporating the supergravity multiplet in the background.
The inclusion of these auxiliary fields eliminated the need to modify the transformation rules of the fields each time a new theory was considered.\footnote{Another reason for using auxiliary fields is that their presence simplifies the couplings of Faddeev-Popov ghost fields, eliminating quartic ghost interactions.}


\section{Tensor calculus}
\label{ss:tensorcalculus}

Having multiplets that transform under a closed algebra makes it possible to combine them in a manner analogous to the sums and products of tensors in geometry. This led to the development of a formal framework for manipulating multiplets, known as 'tensor calculus'.

Using the old minimal set of auxiliary fields, this tensor calculus was developed simultaneously by Ferrara-van Nieuwenhuizen \cite{Ferrara:1978jt,Ferrara:1978wj}
and by Stelle-West \cite{Stelle:1978yr,Stelle:1978wj}.  

In the third and final week of a `Gravitation' summer school in Carg\`{e}se (July 1978) \cite{Levy:1979im}, all of these results were presented. 
This marked the first time that an entire week of a school was dedicated to supergravity. During this meeting,  Bruno Zumino shared his results from the perspective of superspace (see next section).  
The first results on general matter couplings with spontaneous symmetry breaking were presented by Sergio Ferrara and Peter van Nieuwenhuizen. Peter also explained conformal symmetries (see Sec.~\ref{ss:superconform}). Joel Scherk covered the essential elements of higher-dimensional theories. Stanley Deser talked about the dynamics of the gravitino. Dan Freedman addressed field representations of the superalgebras, and Marc Grisaru discussed the anomalies. It was my first exposure to supergravity, and I immediately got a very good overview of the main achievements up to that point.

Later, tensor calculus using the new minimal set of auxiliary fields was also developed \cite{Sohnius:1981ia}. 
The transformations of matter multiplets in these contexts involved one or the other set of auxiliary fields, which made the transformations quite complex.

At the end of the tensor calculus procedure, the action is found using a `density formula' for the multiplets constructed from physical multiplets. The idea is the same as integrating over (parts of) superspace. The density formula, specific for each multiplet, starts from the field of highest mass dimension, but has additional terms that depend on the choice of auxiliary field set. This procedure was later simplified through the use of the superconformal approach.

\section{Superspace}
\label{ss:superspace}
Around the same time that tensor calculus in component language was being developed, a parallel development took place using superspace.
For rigid supersymmetry,  the concept of superspace was introduced in \cite{Volkov:1973ix,Salam:1974yz,Ferrara:1974ac}.
In the early days of supergravity, Wess and Zumino began studying supergravity as a geometry of superspace  \cite{Wess:1977fn} 
building on earlier work in  \cite{Arnowitt:1975xg}.
This approach was further developed in  \cite{Grimm:1977kp,Wess:1978bu} and was presented by Bruno Zumino in the afore mentioned Carg\`{e}se summer school \cite{Zumino:1978hz}.
A constraint-free superfield formalism was found by Warren Siegel and Jim Gates \cite{Siegel:1978mj}. 

We now arrive at the time of the first conference entirely dedicated to supergravity: Stony Brook, September 1979  \cite{VanNieuwenhuizen:1979hm}. 
It was a historic moment for supergravity.\footnote{The conference photo later appeared in many places; I have a large version displayed in my home office.}
The status of superspace supergravity was presented in contributions by Warren Siegel, Julius Wess, Jim Gates, Ulf Lindstr\"{o}m and Martin Ro\v{c}ek.

At that time, supergravity researchers were often divided into two camps: `component' people or `superspace' people, engaged in a friendly competition. This rivalry was evident in paper titles such as
\emph{'The component formalism follows from the superspace formulation of supergravity'}  \cite{Wess:1978ns}  
and, in the reverse direction, the opening line of \cite{Ferrara:1979pk}:  
\emph{'the superspace formalism follows from the component formalism.'}
During the Stony Brook conference, a soccer match was organized to settle which approach was superior: components or superspace. In the end, ordinary space triumphed over superspace with a score of 7–1.
Of course, soon after, these divisions faded as researchers collaborated, integrating both methods. The supergravity community became a group of friends who have achieved remarkable progress over the 50 years of supergravity.

New methods were also presented during the Stony Brook meeting. Notably, the group manifold method (see Sec.~\ref{groupmanifold}), developed in Torino by Tullio Regge's group, was presented by Pietro Fr\'{e}  \cite{DAdda:1979ohp}. However, what struck me most at this meeting was the progress in the development of the supergraph method (also already used in rigid supersymmetry \cite{Grisaru:1979wc}). Mark Grisaru presented his collaborative work with Martin Ro\v{c}ek and Warren Siegel  \cite{Grisaru:1979fy}. At that time, they, along with Jim Gates, were already working on what would later become their standard work in superspace:\footnote{Another major reference on superspace, for which the first version was also published in 1982-83, is the Wess and Bagger  book \cite{Wess:1992cp}. Superspace appeared also in Peter West's 1986 book \cite{West:1986wua}.
 In 1995, another standard work on superspace was written by Buchbinder and Kuzenko \cite{Buchbinder:1998qv}.}
the big book with 1001 lessons: \cite{Gates:1983nr}. 
I also refer to the contribution by Gates, Ro\v{c}ek and Siegel in this half-century book for a more detailed account of supergravity and superspace.

By that time, it seemed that most of supergravity had already been discovered. The Stony Brook proceedings \cite{VanNieuwenhuizen:1979hm} could conclude with a list of 400 papers, which were claimed to be a complete bibliography of supergravity up to that point.

\section{General couplings and supersymmetry breaking}
\label{ss:gencoupl_susybreaking}
The discovery of off-shell formulations made it possible to construct general matter-coupled Lagrangians and initiate a systematic study of the Super-BEH (Brout-Englert-Higgs)  effect, the mass generation of the gravitino as a result of the spontaneous breaking of local supersymmetry.
\subsection{Supersymmetry breaking and super-BEH effect}
Supersymmetry breaking had been extensively studied in rigid supersymmetry, with two main mechanisms: the Fayet-Iliopoulos term
 \cite{Fayet:1974jb}, 
where the auxiliary field of a vector multiplet acquires a vacuum expectation value, and the O'Raifeartaigh model \cite{Fayet:1975ki,ORaifeartaigh:1975nky},
where the auxiliary field of chiral multiplets take a vacuum value. In both cases, the vacuum state has positive energy.

Meanwhile other approaches to supersymmetry breaking had already been explored. The Volkov-Akulov Lagrangian \cite{Volkov:1973ix},  
introduced in 1973, provided an early example of spontaneous supersymmetry breaking with a Goldstone fermion, `Goldstino', making it a key model for future studies of supersymmetry breaking. In their concluding discussion, Volkov and Akulov
mentioned that if a theory of local supersymmetry were discovered, the spin-3/2 gauge field would absorb the Goldstino through the super-BEH effect. This idea was further developed by Volkov and Soroka  \cite{Volkov:1973jd}.
Following the discovery of supergravity, Stanley Deser and Bruno Zumino (April 1977) \cite{Deser:1977uq} 
upgraded this mechanism, demonstrating that spontaneously broken supergravity could be possible with a vanishing cosmological constant.

A concrete model with a vanishing (or, if preferred, a small positive) cosmological constant was found by Pol\'{o}nyi during his stay at Paris ENS \cite{Polonyi:1977pj}. To achieve this, he extended the scalar massive matter models introduced in  \cite{Das:1977mg} by Das, Fischler and Ro\v{c}ek. 
Shortly thereafter, these authors also formulated a broader class of models \cite{Das:1977pu}.
Despite never being formally published, the Pol\'{o}nyi model became a widely used standard example due to its simplicity and illustrative power.
\subsection{The general \texorpdfstring{${\cal N}=1$}{N=1} matter coupling }
\label{ss:N1mattercoupling}
Using the Poincar\'{e} tensor calculus with the old minimal set of auxiliary fields, a large group working at Paris ENS and CERN constructed the general coupling of a single chiral multiplet to supergravity   \cite{Cremmer:1978iv,Cremmer:1978hn}.
These actions included a general kinetic term and a general superpotential term. However, the authors realized that the final form of the action depended only on a single real function ${\cal G}(z,z^*)$ of the complex scalar $z$.
This function was later called the K\"{a}hler-invariant function, but at the time, K\"{a}hler geometry and its associated symmetry had not yet been explicitly introduced (we will revisit this in Sec.~\ref{ss:geometry}).
The results were already presented by Sergio Ferrara in the Carg\`{e}se summer school \cite{Ferrara:1978pb}.
The absorption of the Goldstino by the gravitino was thoroughly discussed. The paper  \cite{Cremmer:1978hn} also established conditions for achieving a minimum of the potential with a vanishing cosmological constant and provided a mass formula, important for realistic models. While the equations were derived in a general framework, the authors primarily focused on `minimal coupling' kinetic terms for the complex scalar: ${\cal L}= -\sqrt{g}\partial _\mu z\,\partial ^\mu z^*$.

I became involved in this endeavor when Peter van Nieuwenhuizen invited me to Stony Brook for two months and posed the challenge of formulating a general action for massive vector multiplets. 
At the same time Sunil Mukhi was given a related problem by Dan Freedman: to construct an action for a massive vector multiplet using the Noether method \cite{Mukhi:1979wc}. 
In my own work \cite{VanProeyen:1979ks}  
I employed the Poincar\'{e} tensor calculus, introducing an arbitrary function $J(C)$ of the real multiplet $C$. Sunil Mukhi's action was later identified as the special case $J=-\ft12 \kappa ^2 m^2 C^2$.
I had the opportunity to present this general coupling of a massive spin $(1,\ft12,\ft12,0)+(2,\ft32)$ multiplet at the Stony Brook meeting  \cite{VanProeyen:1979SBProc}.

When I became a CERN fellow in August 1981, my previous work led Sergio Ferrara to invite me to join their project on formulating the general coupling of chiral and vector multiplets.
This undertaking was, first and foremost, an exercise in tensor calculus. By that time Peter van Nieuwenhuizen had written extensive reviews on the subject, including those for the Carg\`{e}se school \cite{vanNieuwenhuizen:1978st} and his Physics Report \cite{VanNieuwenhuizen:1981ae}, 
which had become the standard reference for supergravity, encapsulating all major developments up to that period.

The outcome of this work was particularly elegant  \cite{Cremmer:1982wb,Cremmer:1982en}. 
It contained the coupling of a set of chiral multiplets to a set of gauge multiplets. The kinetic terms and the scalar potential of the complex scalars $z^\alpha $ (and their supersymmetric partners) are determined by the real function ${\cal G}$, with positive definite metric
\begin{equation}
  {\cal G}_{\alpha \bar \beta }= \partial _\alpha \partial _{\bar \beta }{\cal G}\,.
 \label{galphabarbeta}
\end{equation}
 These scalars and their partners in general transform under the gauge symmetries of the vectors $A_\mu {}^A$ in the vector multiplets, parametrized by $\theta ^A$, e.g. $\delta z^\alpha = \theta ^A k_A{}^\alpha (z)$. These transformations must leave ${\cal G}$ invariant
\begin{align}
  \delta {\cal G}(z,\bar z) & = \theta ^A k_A{}^\alpha\partial _\alpha {\cal G}(z,\bar z) + \hc=0\,,\qquad \mbox{i.e.}\qquad  {\cal P}_A \equiv \rmi \kappa ^{-2}k_A{}^\alpha\partial _\alpha {\cal G} \mbox{ real.}
  \label{PA}
\end{align}
This condition introduces the real function ${\cal P}_A$, known as the moment map.\footnote{These structures are naturally related to K\"{a}hler geometry but that became apparent only later.}

The kinetic term of the vectors is determined by a holomorphic $f_{AB}(z)$  with positive definite real part. Its gauge transformation should satisfy
  \begin{equation}
  k_C{}^\alpha(z)\partial _\alpha  f_{AB}(z)\,=\,2  f_{C(A}{}^D f_{B)D}(z)+\rmi C_{AB,C}\,.
 \label{invffC}
\end{equation}
where $f_{AB}{}^C$ are the structure constants of the gauge algebra.
The possibility of the constants $C_{AB,C}$ was only investigated later, related to possible Chern-Simons terms and/or anomalies \cite{DeRydt:2007vg}.
The moment maps ${\cal P}_A$ must transform in the adjoint representation:
\begin{align}
  k_A{}^\alpha \partial _\alpha {\cal P}_B - k_B{}^\alpha \partial _\alpha {\cal P}_A = f_{AB} {}^C {\cal P}_C\,.
  \label{equivariance}
\end{align}
In the case of abelian symmetries, this condition allows for an additional constant in the moment map, known as the Fayet-Iliopoulos (FI) constant.\footnote{Some of the conditions governing supergravity couplings with a FI terms, particularly in relation to R-symmetry, were already obtained in  \cite{Barbieri:1982ac}.}  

Supersymmetry breaking occurred naturally, with mass terms for observable scalars being easily generated, thus resolving the partner-splitting problem.
General expressions for the Goldstino were derived: it is a linear combination of all the fermions, weighted by the `fermion shifts', the scalar part of their supersymmetry transformation, which correspond to the value of the auxiliary fields.
In rigid supersymmetry,
the Goldstino is massless. In supergravity it is absorbed by a redefinition of the gravitino, which acquires a mass $\kappa ^{-1} < \rme^{{\cal G}/2}>$.
The scalar potential is a square of these fermion shifts, determined by the `metric' defined by the kinetic terms:
\begin{align}
  V= & \kappa ^{-4} \rme^{{\cal G}}\left(-3+\partial _\alpha {\cal G}\,{\cal G}^{\alpha \bar \beta }\,\partial_{\bar \beta } {\cal G}\right)  +\ft12{\cal P}_A\left( \Re f\right) ^{-1\,AB}
{\cal P}_B\,,
  \label{VN1}
\end{align}
where ${\cal G}^{\alpha \bar \beta}$ is the inverse of ${\cal G}_{\alpha \bar \beta }$. For the gravitino $\psi _\mu $, the fermion shift is the part of its transformation proportional to $\gamma _\mu $.
This term gives the negative part of the potential, as the `metric' for that part of the gravitino is negative. This effect is more clearly understood in the conformal formalism.
That negative contribution enables supersymmetry breaking with vanishing cosmological constant.

Meanwhile, also investigations of matter couplings with the new minimal set of auxiliary fields were performed \cite{Sohnius:1982kh}. 

To compare the matter couplings using different sets of auxiliary fields, superconformal tensor calculus (see next section) was employed \cite{Ferrara:1983dh}.  
The results showed that the broadest class of theories could be constructed using the old minimal auxiliary fields.\footnote{This applies to actions with at most two derivatives; for higher-derivative actions, this conclusion may not hold.} Under certain conditions, such as R-symmetry, other sets of auxiliary fields can also be used. In such cases, the results can be transformed into each other though a kind of Legendre transformation with multiplets. It was further demonstrated that replacing chiral multiplets with tensor multiplets does not introduce new physical interactions.

The Lagrangians of matter coupled to ${\cal N}=1$ supergravity  include `soft breaking terms'. These terms,  considered by Luciano Girardello and Mark Grisaru \cite{Girardello:1981wz} 
(and also by Savas Dimopoulos and Howard Georgi   \cite{Dimopoulos:1981zb} 
 in specific models), were introduced into rigid supersymmetric theories to break the symmetry without generating undesirable quadratic divergences. Soft breaking terms were added arbitrarily to the supersymmetric theory without being motivated by any underlying principle. An appealing feature of supergravity matter couplings is that these are automatically generated within the framework of supergravity theories  \cite{Barbieri:1982eh}.  
The mechanism had already been anticipated by \cite{Ovrut:1981zu}.  

\subsection{Phenomenology and no-scale supergravity}
The development of general matter couplings paved the way for constructing realistic particle models with spontaneously broken supersymmetry, triggered by supergravity  \cite{Chamseddine:1982jx,Hall:1983iz}.
In these models, chiral multiplets are present in two sectors: a `hidden sector' at a high mass scale, where supersymmetry breaking occurs, and an observable sector. The two sectors are only coupled through gravitational interactions. As a result, the observable sector only perceives the soft breaking terms, which modify the supertrace formula for masses in rigid supersymmetry \cite{Ferrara:1979wa}. 
In \cite{Cremmer:1982en}, these mass formulas were obtained for standard kinetic terms. Less than two months later, they were generalized to an arbitrary K\"{a}hler model (also using superspace) in \cite{Grisaru:1982sr}.  
Much later, as interest shifted toward de Sitter models, these formulas were extended to curved spacetime in \cite{Ferrara:2016ntj}.
A thorough review of the phenomenological applications explored during the early years can be found in \cite{Nilles:1983ge}. 

A particular class of models that deserved significant attention were the so-called `no-scale supergravities'.
It was shown in \cite{Cremmer:1983bf} 
that ${\cal N}=1$ supergravity can be spontaneously broken in sectors with scalars that possess a flat direction in the potential (and therefore often referred to as `flat potential' models). In these models, the supergravity breaking scale (and the gravitino mass) is arbitrary at tree level. This subclass of spontaneously broken supergravity models is also known as models with `naturally vanishing cosmological constant'.
Some of their key properties were already identified in \cite{Chang:1983hk}. 
The mechanisms behind no-scale supergravities were soon applied in supersymmetric model building in \cite{Ellis:1983sf,Ellis:1984bm}. 
Ed Witten \cite{Witten:1985xb} soon thereafter discovered that such models emerge in the compactification of superstring theories on a Calabi-Yau manifold. 
The contribution of Fabio Zwirner in this book will provide further insight into these `no-scale models'.

\section{Superconformal methods}
\label{ss:superconform}

A framework that unifies the various Poincar\'{e} tensor calculi is the superconformal method.
The superconformal algebra was identified by Haag, {\L}opusza\'nski, and Sohnius \cite{Haag:1974qh} as the unique extension of the super-Poincar\'{e} algebra.  
Under the condition that the bosonic subalgebra is a direct sum of the conformal algebra and R-symmetries, Werner Nahm \cite{Nahm:1978tg} 
further classified the finite-dimensional superconformal algebras, which exist only in dimensions up to six.\footnote{Nahm’s restrictions could be circumvented by a soft algebra with structure functions rather than structure constants, as e.g. in \cite{Bergshoeff:1982az} and more recently in  \cite{Adhikari:2023tzi}. However, such theories lack a vacuum that preserves conformal symmetry because the dilaton, which transforms under dilatations, is always nonzero. Consequently, this soft algebra has no rigid counterpart.}

The conformal algebra had already appeared in the first field-theoretic model of supersymmetry by Julius Wess and Bruno Zumino  \cite{Wess:1974tw}. 
Using this algebra, curvatures for the superconformal generators were constructed in  \cite{Kaku:1977pa,Ferrara:1977ij}, where the authors explored superconformal actions quadratic in the curvatures.
The key insight—using superconformal symmetry as a tool to construct super-Poincar\'{e} actions—was introduced in 1978 by Michio Kaku and Paul Townsend \cite{Kaku:1978ea}, who demonstrated how Poincar\'{e} supergravity could be obtained as a broken phase of superconformal gravity.
Kaku presented these results at the Stony Brook conference \cite{Kaku:1979gc} and explained how the auxiliary fields for Poincar\'{e} supergravity emerge from this procedure.


The superconformal method uses the procedure of `gauge equivalence', meaning that, during the construction of a theory, one initially introduces more symmetry than is required in the final model. The additional symmetry is not meant to be a fundamental feature of the theory but rather serves as a \emph{tool} to systematically derive all super-Poincar\'{e} invariant actions. While we do not enforce superconformal symmetry in the final action, its presence during construction greatly simplifies the formulation of supergravity theories.

First one defines the \emph{Weyl multiplet}, i.e. a multiplet that contains all the gauge fields associated with the superconformal algebra. This multiplet was found for ${\cal N}=1$ in 1977, in the context of superconformal actions quadratic in curvatures \cite{Kaku:1977rk,Kaku:1978nz,Townsend:1979ki}. Then the matter multiplets must be adapted to transform consistently under local superconformal transformations, using the Weyl multiplet as a background.

One (or for extended supergravity more) of these multiplets must be selected as  compensator. That multiplet should contain fields to `compensate' for the additional symmetry that is in the superconformal algebra. The choice of compensator multiplet determines the auxiliary field formulation that emerges in the final super-Poincar\'{e} theory. E.g. using a chiral multiplet leads to the old minimal formulation. Using a linear multiplet leads to the new minimal formulation, and using a complex multiplet leads to the non-minimal (Breitenlohner) set of auxiliary fields.

Superconformal calculus enables the construction of actions through `density formulas', defining an invariant action from specific combinations of fields in composite multiplets. This is similar to what we mentioned at the end of Sec.~\ref{ss:tensorcalculus} for tensor calculus in general, but here it is a formula that gives a superconformal invariant action.

Then one has to choose gauge fixings for the extra symmetries that are in the superconformal algebra and not in the super-Poincar\'{e} algebra (dilatations, special conformal transformations, the R-symmetry $\U(1)$ and the S-supersymmetry). In that way the symmetry is reduced to super-Poincar\'{e}. In the original papers, gauge fixing was performed by setting a field to a constant or zero. Improved gauge conditions were found in \cite{Kugo:1982mr}. Before this, in  \cite{Cremmer:1978hn,Cremmer:1982en}, cumbersome field redefinitions were required to obtain kinetic terms properly separating the various physical fields. A typical example is a suitable gauge choice such that the scalar curvature appears as in the Einstein-Hilbert action multiplied with $(2\kappa ^2)^{-1}$ (`Einstein frame'), eliminating the need for later field redefinitions.

Finally, one rewrites the action in terms of a parametrization solving the gauge conditions and identifying combinations of the superconformal transformations that preserve the gauge fixing conditions. For instance, the supersymmetry that survives in the final Poincar\'{e} theory is a field-dependent combination of the $Q$-supersymmetry, $S$-supersymmetry, dilations, $R$-symmetry and special conformal symmetry of the superconformal algebra. These rules for defining the Poincar\'{e} transformations in terms of the conformal ones are known as `decomposition laws'.

A more complete calculus with all types of multiplets was later found in \cite{Kugo:1983mv}. 

Superconformal tensor calculus provides a general and flexible framework for studying matter interactions without requiring separate tensor calculi. Its usefulness is particularly evident when the theory is more complicated. I came to appreciate its power especially in the context of ${\cal N}=2$.

Conformal supergravity has also been studied in superspace   \cite{Gates:1979yz,vanNieuwenhuizen:1979xf,Howe:1981gz} 
and in the group manifold approach \cite{Castellani:1981um}. 
Relations between the conformal supergravity in ordinary space and its superspace constraints have been found in \cite{vanNieuwenhuizen:1979xf}. 

Using conformal methods (along with geometry treated in Sec.~\ref{ss:geometry}), the paper \cite{Cremmer:1982en}, discussed a lot in Sec.~\ref{ss:gencoupl_susybreaking}, was for a large part rewritten with cosmological applications in mind in \cite{Kallosh:2000ve}. A pedagogical exposition of these methods can be found in \cite{Freedman:2012zz}.

\section{Group manifold approach}
\label{groupmanifold}
I first encountered this third method for constructing supergravity theories during the Stony Brook workshop  \cite{DAdda:1979ohp}.  Known as the geometric or rheonomic approach, it was pioneered by Ne'eman and Regge \cite{Neeman:1978njh}, 
 and further developed by the group in Torino  \cite{DAdda:1980axn,DAuria:1980cmy}. 
Our understanding of the approach deepened after a seminar\footnote{It was still at the time that slides had to be written by hand. Tullio, being surprised that he was scheduled to lecture early in the meeting, used his `students' Riccardo D'Auria and Pietro Fr\'{e} to write the slides while he was already speaking. This led to some confusion in the order of the slides that they put in front of Tullio. Difficult to tell to those who were not present, but an amusing and unforgettable story for the participants.}
 by Tullio Regge in Erice, March 1980 \cite{DAuria:1980bln}.

The method begins with a supergroup, treating fields as components of one-forms on the corresponding supergroup manifold.\footnote{As such, it can be viewed as an extension of superspace where fields appear as components of a superfield.}
Transformations in this formalism are interpreted as superdiffeomorphisms on the supergroup manifold, which lead to a `Free Differential Algebra' \cite{D'Auria:1982nx} associated with the super-Poincar\'{e} group. 
 Spacetime emerges as a bosonic embedding within the super-Poincar\'{e} manifold.
`Horizontality conditions' are imposed, ensuring that curvature components vanish along the direction of that spacetime manifold. These conditions are essential for interpreting the diffeomorphisms as a spacetime symmetry. 
The remaining transformations are the `rheonomic symmetries'. Gauge transformations correspond to diffeomorphisms along the directions where curvatures remain horizontal.
The Bianchi identities 
define a soft group manifold. The action is then constructed as an integral of a $D$-form built from gauge fields and field strengths on that soft group manifold. Finally, the coefficients in the action are fixed by requiring the existence of a vacuum solution: imposing that setting curvatures to zero must satisfy the equations of motion.

Through this approach, pure supergravity theories were obtained, e.g. in $D=4$ in  \cite{DAuria:1980cmy}  
and in $D=5$ in  \cite{DAuria:1981yvr}. 
Matter couplings were incorporated   \cite{DAuria:1980ffr} 
by extending the formalism with 0-forms. In this framework, certain
curvature components were identified with the auxiliary fields, obtaining in that way the old minimal set of auxiliary fields.
The method also allowed the construction of theories lacking an invariant action, such as in $D=6$ in \cite{DAuria:1983jkr}. 

The workshop in Karpacz, Poland, February 1983 \cite{Milewski:1985qr}, was particularly useful. It provided a detailed review and comparison of the group manifold approach \cite{Castellani:1985ns}, superconformal tensor calculus \cite{VanProeyen:1983wk} and superspace methods \cite{Galperin:1985nt}, together with contributions on e.g. extended supersymmetry and geometry. It was a meeting where many supergravity practitioners joined and collaborated.

The group manifold approach will be discussed in greater depth in this book by Riccardo D'Auria, Leonardo Castellani and Pietro Fr\'{e}, who also authored the standard work on the subject in 1991 \cite{Castellani:1991books123}.
For more recent and shorter reviews, see  \cite{Castellani:2022iib,Andrianopoli:2024qwm}.

\section{K\"{a}hler geometry}
\label{ss:geometry}
When discussing matter couplings today, one typically starts from K\"{a}hler geometry. However, while preparing this review, I was reminded that these couplings were originally developed without explicit reference to this geometric framework. At the time, our primary focus was on `minimal coupling' kinetic terms, which, in K\"{a}hler terminology, correspond to a flat metric $g_{\alpha \bar \beta }= \delta _{\alpha \beta }$. Indeed, that seemed to be the most relevant one for supersymmetric phenomenology.

Zumino in August 1979 \cite{Zumino:1979et} 
supersymmetrized (rigid supersymmetric) nonlinear sigma models with scalar fields taking values in a K\"{a}hler manifold.
Rather than deriving K\"{a}hler geometry as a consequence of chiral multiplet couplings, as we often describe it today, Zumino started from K\"{a}hler geometry itself. His result is now recognized as the general kinetic action for chiral multiplets.

We arrive now to the time of the Erice conference, March 1980.
In my memory, it was the third important meeting \cite{Ferrara:1980nr} (after the mentioned school in Carg\`{e}se and the Stony Brook workshop.)\footnote{However, my mandatory service in the Belgian army prevented me from attending another highly significant event: the Cambridge meeting in the summer of 1980 \cite{Hawking:1981bu}.
In the following years, the Trieste spring schools and workshops became the main gathering points for the supergravity community, holding a status comparable to today’s annual  String conferences. Abdus Salam enthusiastically supported these events.  The first one took place in the spring 1981 \cite{Ferrara:1982be}, and despite my military obligations, I managed to attend—albeit by illegally leaving the country during my service period.  }
Unfortunately, this was the last time that we met Joel Scherk.  In recognition of his many groundbreaking contributions to Superstrings and Supergravity, the book was later dedicated to his memory.

In that Erice workshop, Luis Alvarez-Gaum\'{e} and Daniel Freedman  \cite{Alvarez-Gaume:1980xas} have explained K\"{a}hler geometry. At the time, this topic was not widely known among physicists, although Jerzy Lukierski had already discussed K\"{a}hler and quaternionic (super)geometry at the Stony Brook conference \cite{Lukierski:1979bf}. However, its significance was not fully appreciated at the time. The introduction by Alvarez-Gaum\'{e} and Freedman, published in the Erice proceedings, remains a nice introduction to K\"{a}hler geometry.
Later, in  \cite{Alvarez-Gaume:1981exv} 
they also explained that hyper-K\"{a}hler manifolds appear in extended supersymmetry (as discussed further in Sec.~\ref{ss:quatSpecGeom}).

The crucial role of K\"{a}hler geometry became evident after the Bagger-Witten papers  \cite{Witten:1982hu,Bagger:1982ab}.
The first has the title `Quantization of the Newton's constant', which is a quantization in terms of the scalar self coupling in presence of non-contractible 2-cycles in the manifold. The K\"{a}hler manifolds that satisfy these global restrictions are the K\"{a}hler-Hodge manifolds.

Today, the geometry of the scalar manifold provides the framework to classify matter couplings, and its importance grows even further in extended supergravity.
 In Sec.~\ref{ss:N1mattercoupling}, the key aspect of matter couplings is now always framed in terms of `choosing a K\"{a}hler manifold' (or more precisely, a K\"{a}hler-Hodge manifold), with a K\"{a}hler potential ${\cal K}(z,\bar z)$, and a superpotential $W(z)$. The function ${\cal G}(z,\bar z)$, mentioned in Sec.~\ref{ss:N1mattercoupling}, is related to these quantities as follows:
\begin{equation}
  {\cal G}(z,\bar z)= \kappa ^2{\cal K}(z,\bar z) +\log[ \kappa ^6 W(z)\bar W(\bar z)]\,.
 \label{calGinKW}
\end{equation}

\section{\texorpdfstring{${\cal N}=2$}{N=2}.}
\label{ss:N2}
\subsection{The first models and auxiliary fields}
We now go back in time. The first ${\cal N}=2$ supergravity models, at the time only for the gravity multiplet with spins $(2,\ft32,\ft32,1)$, were constructed using the Noether method   \cite{Ferrara:1976fu,Freedman:1976aw,Ferrara:1976iq}
and combining already known ${\cal N}=1$ theories.

The earliest ${\cal N}=2$ matter coupling I am aware of is found in  \cite{Luciani:1977hp},   
where multiple vector multiplets (spin $(1,\ft12,\ft12,0,0)$) were coupled to supergravity. The other fundamental physical multiplet is the one with spins $(\ft12,\ft12,0,0,0,0)$, now commonly referred to as `the hypermultiplet'. Interestingly, Pierre Fayet originally used the term `hypersymmetry' to describe ${\cal N}=2$ supersymmetry  \cite{Fayet:1975yi} and he also used `hypermultiplets' for what we denote now as vector multiplets. He discussed the different mechanisms for generating mass \cite{Fayet:1978ig}, now known as the Higgs and Coulomb branches. Later, the term hypermultiplet became exclusively associated with the multiplet containing spins $(\ft12,\ft12,0,0,0,0)$.

 The first coupling of the hypermultiplet to supergravity appeared in  \cite{Zachos:1978iw}.  
A crucial feature of the massive hypermultiplet is that it requires central charges in the supersymmetry algebra, which in supergravity are gauged by the `graviphoton', the spin 1 particle in the gravity multiplet. In his Ph.D. thesis, Cosmas Zachos   \cite{Zachos:1979uh}, 
pointed out that this mechanism results in a negative contribution to the gravitational attraction between massive objects. This phenomenon was humorously dubbed `antigravity' by Joel Scherk in his paper “Antigravity: A crazy idea?"  \cite{Scherk:1979aj}. 
He discussed this in the Stony Brook conference \cite{Scherk:1979rh} and in his last seminar in Erice  \cite{Scherk:1980gq} he  sketched experimental setups to detect deviations in gravity. Interesting, this `crazy idea' later proved to be the essential ingredient in the force cancelation between  parallel BPS D-branes.

Peter Breitenlohner\footnote{The paper was published with the authors listed as Breitenlohner and A. Kabelschacht. According to Martin Sohnius : “Well before 1975, there used to be a routine turf war between traditionalists and modernisers whether to put academic titles on the name plates on people's office doors.  Apparently, this went back and forth a few times, and on one occasion when titles were added, some junior housekeeping person had taken all the name plates down and wrote up a list of what the new ones, with titles, should be.  He came across a plate that said "Kabelschacht" and had been taken off a door to a service duct ("cable shaft").  He checked the staff roster, was puzzled, and turned to the head porter, Herr Cirpka, who was himself part of the furniture at the MPI, having come with Heisenberg from G\"{o}ttingen when the Institute was moved to Munich in the early 60's.  What is the proper title for Kabelschacht?  Cirpka gave the dead-pan answer "Professor, of course," and thus the door was labelled "Prof. Kabelschacht".  The door and sign remained as an in-house joke at the MPI.
Years later, Breitenlohner decided that it was "about time that Prof. Kabelschacht actually had a publication list," gave him the first name Alois, which to a German rings of a very old fashioned Bavarian in lederhosen, smoking a pipe with a lid on it and wearing a hat with a goose feather, and put him as a second author on a paper which he thought to be fairly unimportant.”
}
 \cite{Breitenlohner:1978rq} obtained auxiliary fields for $N=2$ in 5 and 6 dimensions. 

Soon afterward, auxiliary fields for ${\cal N}=2$ supergravity in $D=4$ with linearized transformations were independently obtained by Fradkin and Vasiliev
 \cite{Fradkin:1979as,Fradkin:1979cw} 
and by de Wit and van Holten \cite{deWit:1979pq}. 
These multiplets have 40+40 components, constructed by combining ${\cal N}=1$ multiplets with an off-shell spin $(\ft32,1)$ multiplet. An earlier version of such a multiplet had been introduced in
\cite{Ogievetsky:1975vk,Ogievetsky:1976qb} 
as a gauge spinor superfield, but  \cite{Fradkin:1979as,deWit:1979pq,Fradkin:1979cw} used an alternative formulation. These differences were later analyzed in  \cite{Gates:1979gv}. 

This occurred during my visit to Stony Brook mentioned above, where Bernard de Wit was also visiting. In this way, I began working on ${\cal N}=2$ supergravity, without realizing that it would become a main part of my research for the next 30 years. After returning to Leuven, our collaboration continued, now also involving his Ph.D. student Jan-Willem van Holten. This led to many car rides between Leiden, Amsterdam and Leuven, marking the beginning of a long and productive research journey.

Meanwhile, Breitenlohner (now collaborating with Martin Sohnius) also obtained the auxiliary fields, and started tensor calculus for ${\cal N}=2$ \cite{Breitenlohner:1979np}. 
At the Stony Brook conference, both Bernard de Wit and Peter Breitenlohner presented their results on ${\cal N}=2$ auxiliary fields and tensor calculus.

We had obtained the full transformation rules and action \cite{deWit:1980ug} 
and obtained the superconformal Weyl multiplet, the starting point for the superconformal tensor calculus of ${\cal N}=2$. We realized that the super-Poincar\'{e} theory with auxiliary fields could be decomposed into a 24+24 Weyl multiplet and two 8+8 multiplets: a vector multiplet (containing the graviphoton) and a non-linear second compensating multiplet.

Soon after, we obtained the local superconformal formulation of the chiral multiplet \cite{deRoo:1980mm} 
with its density formula. This would later lead to a deeper understanding of vector multiplet couplings (see Sec.~\ref{ss:quatSpecGeom}).

We are now also back to the time of the Erice conference, march 1980 \cite{Ferrara:1980nr}. It was the occasion to learn from discussions with Peter Breitenlohner on their approach to ${\cal N}=2$, and combining possible couplings of the physical ${\cal N}=2$ multiplets (vector multiplet and hypermultiplet) to supergravity.
To keep discussions lively, even at night, Peter Breitenlohner and Bernard de Wit were assigned to the same room at the Erice conference. With the help of some Sicilian drinks, differences in their approaches could be thoroughly debated.

The hypermultiplet was the most tricky one. It can be defined off-shell only in presence of central charges. These features eventually led to the development of harmonic superspace, which will be reviewed in Sec.~\ref{ss:harmonic}.
 In  \cite{deWit:1980gt} 
it was shown how such a (field-dependent) central charge could be defined within the superconformal framework. The fields of a compensating vector multiplet then enter in the `background' of the hypermultiplet, just as the Weyl multiplet serves as the background for all matter multiplets. Its gauge field (which becomes the graviphoton in the Poincar\'{e} theory) gauges the central charge, and the central charge transformed fields of hypermultiplet’s four scalar fields act as auxiliary fields in the action. Using this hypermultiplet as a second compensating multiplet (in addition to the vector multiplet) led to a second set of 40+40 off-shell fields for ${\cal N}=2$ supergravity as was at the end of that year clarified in  \cite{deWit:1981tn}, 
where the structure of the conformal calculus was fully exhibited.

Meanwhile, significant progress was made also using superspace methods \cite{Breitenlohner:1980ej,Howe:1980sy}, and, as is often the case, these developments stimulated each other.
Breitenlohner and Sohnius obtained also the coupling of hypermultiplets to supergravity \cite{Breitenlohner:1981sm}, but the results in  \cite{deWit:1980gt,Breitenlohner:1981sm} were limited to projective quaternionic spaces and did not provide a fully general coupling for hypermultiplets.

An investigation on a superconformal version of the tensor multiplet \cite{deWit:1982na}  
led to a third version of a 40+40 fields for off-shell ${\cal N}=2$ Poincar\'{e} gravity. Bernard de Wit presented an overview at the first String meeting in Trieste
 \cite{deWit:1981vgr}. 

For rigid supersymmetry, another off-shell formulation of the hypermultiplet, the so-called `relaxed hypermultiplet'  \cite{Howe:1982tm} 
was found with unconstrained superfields. However, as far as I know, this was never coupled to supergravity. The hyper-K\"{a}hler geometry described corresponds to the rigid limit of those in \cite{deWit:1980gt,Breitenlohner:1981sm}.

\subsection{The general theory}

With sufficient groundwork laid, it became possible to explore general couplings systematically. The initial step was to construct models that were quadratic in the fields before coupling them to supergravity \cite{deWit:1984rz}. 
After eliminating the auxiliary fields, the resulting supergravity-coupled models naturally led to nonlinear sigma models, with the vector multiplet scalars parameterizing a complex projective space and the hypermultiplet scalars a quaternionic projective space. The structure of Yang–Mills theories, including the induced scalar potential, was analyzed in detail \cite{Derendinger:1983rc} 
and compared to the ${\cal N}=1$ case \cite{Derendinger:1984gf}.

It was soon realized that more general models are possible based on a holomorphic function $F(z)$ of the scalars \cite{deWit:1984pk}.  
These were studied also in the context of duality transformations of the vectors, eventually giving rise to what later became known as `special K\"{a}hler manifolds'.
A major European collaboration explored their geometric properties, gaugings, scalar potentials, and the super-Higgs effect  \cite{Cremmer:1985hj}. 
The structure of potentials in ${\cal N}=2$ supergravity-matter couplings was examined, drawing comparisons with the ${\cal N}=1$ case and refining the understanding of the super-BEH effect.

These advancements were then extended to include hypermultiplets containing fields that transform under the gauge group of the vector multiplets. The resulting general action was further enriched by the inclusion of a Chern-Simons term. This paper contained the general matter couplings, encompassing all ${\cal N}=2$ previously constructed  theories   \cite{deWit:1985px}. 

\subsection{Harmonic Superspace}
\label{ss:harmonic}
As previously mentioned, ${\cal N}=2$ supergravity has also been studied using superspace. The central charges, especially for hypermultiplets, complicates matters. The development of `harmonic superspace' provided a way to address these issues. This approach extends conventional superspace by introducing additional bosonic coordinates that describe the sphere $\SU(2)/\U(1)$. The formalism was primarily developed by a Russian group  \cite{Galperin:1984av}. 
Shortly after its introduction, a related framework known as `projective superspace' was formulated by Anders Karlhede, Ulf Lindstr\"{o}m, and Martin Ro\v{c}ek \cite{Karlhede:1984vr}.  
They used tensor multiplets and the method involved integration over a projection of superspace using also the sphere coordinates.
It was soon applied also to supersymmetry in $D=2$ \cite{Gates:1984nk}, 
leading to `twisted multiplets'.

These methods enabled the construction of a broad class of hypermultiplet couplings, and in the following years, their application extended to ${\cal N}=2$ supergravity. By the end of the period covered in this review, the most general matter couplings had been obtained, culminating in a new formulation of ${\cal N} = 2$ Einstein supergravity featuring an off-shell complex hypermultiplet with infinitely many auxiliary fields \cite{Galperin:1987em,Galperin:1987ek}. 

Harmonic superspace is reviewed in the book  \cite{Galperin:2001uw}  
and in \cite{Kuzenko:2009zu}. 

\subsection{Partial symmetry breaking}
In 1984, Sergio Cecotti, Luciano Girardello, and Massimo Porrati established a theorem \cite{Cecotti:1984rk} 
stating that, within the framework of standard tensor calculus, a partial breaking of supersymmetry from ${\cal N}=2$ to ${\cal N}=1$ is not possible. However, they later demonstrated that by employing duality transformations and Fayet-Iliopoulos terms
\cite{Cecotti:1984fn,Cecotti:1984wn} 
an exceptional model allowing such partial breaking could be constructed \cite{Cecotti:1985sf}.  
Subsequent research expanded on this idea, showing that by incorporating magnetic Fayet–Iliopoulos terms, additional possibilities for partial supersymmetry breaking could be realized. For a summary of the current understanding and recent developments in this area, see the introduction of \cite{Gold:2024gsj}.

\section{Quaternionic and special geometry}
\label{ss:quatSpecGeom}
The couplings of chiral multiplets in ${\cal N}=1$ respect a complex structure $J$ satisfying $J^2=-\unity $, which remains covariantly constant under connections compatible with a Hermitian metric. This naturally defines K\"{a}hler manifolds. The hypermultiplet scalars of ${\cal N}=2$ preserve three such complex structures $J^A$, satisfying the quaternion algebra $J^AJ^B = -\delta ^{AB}\unity +\varepsilon^{ABC}J^C$. For rigid supersymmetry, these structures are covariantly constant, defining hyper-K\"{a}hler manifolds. In supergravity, they are preserved up to local rotations between them, leading to quaternionic-K\"{a}hler manifolds. The canonical paper that explained this for physicists was written by John Bagger and Ed Witten \cite{Bagger:1983tt}. 
A more detailed analysis of quaternionic manifolds followed, and a comprehensive review can be found in this book by Stefan Vandoren \cite{Vandoren:2025izf}.

The vector multiplet scalars in ${\cal N}=2$ are complex, and also define a K\"{a}hler manifold. However, when constructing general vector multiplet couplings, it became evident that not all K\"{a}hler manifolds appear. As previously mentioned, in the tensor calculus the models are built from a holomorphic $F(z)$ rather than a general K\"{a}hler potential ${\cal K}(z,\bar z)$. This restriction led to the identification of a special subclass of K\"{a}hler manifolds governing the complex scalars in ${\cal N}=2$ vector multiplets coupled to supergravity \cite{deWit:1984pk,Cremmer:1985hj}. Over time, geometric characterizations emerged to define that geometry, which was later baptized `special K\"{a}hler geometry' by Strominger \cite{Strominger:1990pd}. 
The defining structure of these manifolds is intimately linked to the electric-magnetic duality of vector fields, which, via supersymmetry, imposes constraints on the scalar sector.

During my postdoctoral stay in Paris with Eug\`{e}ne Cremmer, who had profound mathematical expertise, I had the opportunity to explore the geometric structure of these spaces in detail. Our work led to a classification of symmetric spaces within this class and a derivation of key properties of their curvature tensors~\cite{Cremmer:1985hc}. 

A crucial conclusion from these developments is that the scalar manifold in ${\cal N}=2$ supergravity
takes the form of a product:
\begin{equation}
  {\cal M}_{\rm SK}\times Q\,,
 \label{MSKtimesQ}
\end{equation}
where ${\cal M}_{\rm SK}$ is a special K\"{a}hler manifold and $Q$ a quaternionic-K\"{a}hler manifold.

Around the same time, Philip Candelas, Gary Horowitz, Andy Strominger and Ed Witten  \cite{Candelas:1985en} 
demonstrated in the context of string theory that Calabi-Yau compactifications of $D=10$ supergravities to $D=4$ lead to the supergravity-matter couplings discussed here. In particular, the geometry of the deformations (`moduli') of Calabi-Yau manifolds for compactifications of type $IIA$ and $IIB$ supergravities exhibits the structure (\ref{MSKtimesQ}), further stimulating research into ${\cal N}=2$ matter couplings.

A key development was the c-map  \cite{Cecotti:1989qn}, 
which establishes a correspondence between special K\"{a}hler manifolds and quaternionic-K\"{a}hler manifolds. Additionally, systematic classifications of homogeneous manifolds were pursued \cite{deWit:1991nm}. 
These supergravity constructions also attracted significant interest from the mathematical community   \cite{QuatWorksh1,QuatWorksh2}, 
 since the supergravity constructions led to new results on these manifolds.

A major contribution to this area came from the group manifold approach, which provided a coordinate-independent formulation of special geometries  \cite{Castellani:1990tp,Castellani:1990zd,Andrianopoli:1996vr,Andrianopoli:1997cm}. 
 A nice book on the geometric concepts is  \cite{fresoriabook}, while the tensor calculus results are compiled in \cite{Lauria:2020rhc}.

\section{More supersymmetries and dimensions}
\label{ss:N4D56}
\subsection{\texorpdfstring{${\cal N}=4,\, D=4$}{N=4, D=4}}
\label{ss:N4}
A key step in understanding matter couplings for ${\cal N}=4$ supergravity was taken in \cite{Bergshoeff:1981is,Bergshoeff:1980sw}, where the Weyl multiplet, a 128+128 multiplet, was identified. 
Around the same time, Ali Chamseddine derived a coupling to six vector multiplets via dimensional reduction from $D=10$ \cite{Chamseddine:1980cp}. 
Further advancements came from Paul Howe, who constructed auxiliary fields for linearized ${\cal N}=4$ supergravity (208+208 components)  \cite{Howe:1981hd}. 
He developed a suggestion in   \cite{Howe:1981nz}, 
combining a `minimal' 128+128 multiplet, coupling to the supercurrents, with 5 abelian Yang-Mills  multiplets,
the latter having an off-shell central charge that vanishes on-shell as established in \cite{Sohnius:1980it}. 
The resulting Lagrangian described supergravity coupled to Yang-Mills fields, but it was not possible to separate the two systems off-shell.
Additional progress using the superspace approach was  obtained in  \cite{Gates:1982an}. 
This led also to theories with various scalar potentials \cite{Gates:1982ct,Gates:1983ha}. 

A systematic investigation using the conformal approach was carried out by Mees de Roo  \cite{deRoo:1984gd}, 
who successfully obtained couplings to an arbitrary number ($n$) of vector multiplets. This was achieved by combining the Weyl multiplet with six compensating vector multiplets, leading to a scalar field space described by the coset structure $[\SO(n,6)/(\SO(n)\times \SO(6))]\times [\SU(1,1)/\U(1)]$.
Initially, this was formulated for abelian vector multiplets and later extended to non-abelian couplings  \cite{deRoo:1985np}, 
including cases with non-compact gauge algebras. Further developments, including an analysis of the super-BEH effect with a vanishing cosmological constant, were presented in   \cite{deRoo:1985jh}. 

Matter couplings for ${\cal N}=3$ supergravity were constructed in  \cite{Castellani:1985ka} and presented \cite{Castellani:1985ic}.
at the Torino meeting  \cite{DAuria:1986ysy},  
which marked the conclusion of the period under review.

\subsection{\texorpdfstring{$D=5$}{D=5}}

The matter couplings for ${\cal N}=2$ in four dimensions share similarities with those in $D=5$ with minimal supersymmetry (often referred to as either ${\cal N}=2$ or, in some conventions, ${\cal N}=1$) and in $D=6$ (denoted as either ${\cal N}=2$ or $(1,0)$ due to the chirality of the supersymmetry parameters).
The first constructions of $D=5$ supergravity theories were presented by Eug\`{e}ne Cremmer at the 1980 Cambridge meeting \cite{Cremmer:1980gs}, followed shortly after by work from Ali Chamseddine and Hermann Nicolai  \cite{Chamseddine:1980mpx}. 
Paul Howe constructed superspace multiplets  from supercurrents with appropriate constraints   \cite{Howe:1981ev}. 

In July 1983, I received a letter\footnote{We did not have email yet. They wrote to me to include their work in an overview of supergravity-matter couplings at the Europhysics High Energy conference of 1983  \cite{VanProeyen:1983tx}. } from Paris, written on July 14 by the collaboration of Murat G\"{u}naydin, German Sierra and Paul Townsend. The letter, titled “Bastille Day; Le Jour de Gloire est Arriv\'{e}", described their progress in constructing vector multiplets couplings in $D=5$, leading to an elegant geometric structure connected to the Magic square  \cite{Gunaydin:1983rk}. 
Their work revealed deep connections between matter couplings and Jordan algebras of $3\times 3$ hermitian matrices over the division algebras $\mathbb{R},\, \mathbb{C}, \,\mathbb{H}$ and $\mathbb{O}$. Unital Jordan algebras have a cubic norm form $N=C_{IJK}X^I X^JX^K$, where the structure constants $C_{IJK}$
play a crucial role in both the scalar field geometry and in the vector sector 
 through a Chern-Simons-like term.
This geometric structure became known as `very special real geometry'.
G\"{u}naydin, Sierra, and Townsend continued their investigation in  \cite{Gunaydin:1984bi,Gunaydin:1985ak}, where they systematically analyzed $D=5$ Maxwell-Einstein supergravity theories. Their results established a general framework for theories involving vector and tensor multiplets, all unified by the same Chern-Simons structure. 
I vividly recall being fascinated and inspired by their discoveries. Their work significantly influenced my subsequent research, ultimately leading to our own contributions on general matter couplings in $D=4$ supergravity \cite{deWit:1984rz,deWit:1984pk}.

\subsection{\texorpdfstring{$D=6$}{D=6}}

The pure $N = 2$ supergravity theory in $D = 6$ faces a challenge in constructing a manifestly Lorentz invariant Lagrangian because of the presence of a
self-dual third-rank field strength in the graviton multiplet \cite{Marcus:1982yu}. In order to evade this difficulty, one can couple an
antisymmetric tensor multiplet containing an anti-self-dual field strength to supergravity. Together, these components form a complete antisymmetric tensor, allowing for the construction of a Lorentz-invariant action. In  \cite{Nishino:1984gk}, Nishino and Sezgin 
constructed the couplings of a single antisymmetric tensor multiplet, a Yang-Mills multiplet and hypermultiplets. Abdus Salam and Ergin Sezgin related the $D=6$ Einstein-Maxwell theories to $D=4$ by compactification on a 2-sphere \cite{Salam:1984cj}.  

The superconformal formulation of the $D=6$ theory was subsequently developed in \cite{Bergshoeff:1986mz} and presented at the 1985 Torino meeting mentioned above  \cite{VanProeyen:1985ib,Bergshoeff:1985mv}.

In the years that followed, the relations between very special real, quaternionic, and special K\"{a}hler manifolds were studied in depth. These investigations ultimately led to the development of a general framework now known as special geometries.

\section{Conclusions}
\label{ss:conclusions}
From this review, it is clear that there are many facets of matter couplings. But there are certainly many more important aspects that could not be addressed here. One such topic, especially important for broken supersymmetry, is the method of constrained superfields or multiplets, which has been extensively developed in Padova. This approach led to the definition of pure de Sitter supergravity in \cite{Bergshoeff:2015tra}.
The contribution by Ignatios Antoniadis, Emilian Dudas, Fotis Farakos, and Augusto Sagnotti \cite{Antoniadis:2024hvw} in this book further explores this, with applications to inflationary cosmology.

Matter couplings also emerge from string theory, particularly in orbifold compactification, intersecting branes and flux compactifications  \cite{Giddings:2001yu}.
In the 21st century, the study of supergravity matter has gained momentum, especially in its applications to cosmology, particularly the inflationary universe and its de Sitter phase, in line with the observational results from the Planck mission, see the contribution by Renata Kallosh and Andrei Linde.

Reflecting on the journey of supergravity-matter couplings, I was struck by how much had already been discovered in the early years. But still, at the time we believed we had already obtained a solid understanding of the matter couplings. However, it soon became evident that many additional aspects, such as their intricate geometry, symmetries, dualities, \ldots,  remained to be discovered after 1985, the cutoff date of this review.

For a period, there was considerable hope that the LHC might already measure supergravity matter couplings. This optimism, however, was based on an assumption of a low-energy scale of supersymmetry that was not well founded. The gravitational coupling constant continues to be a fundamental feature that cannot be ignored. But the study of matter couplings will be important. Such a beautiful theory will undoubtedly play an important role in our understanding of the universe.

\subsection*{Note added}
This review concerns the matter couplings of supergravity, which was of course not independent of earlier developments. This has recently been reviewed in \cite{West:2025oji}. In particular the role of  \cite{Chamseddine:1976bf} was important for understanding the gauging of superalgebras as e.g. used in the superconformal approach, and providing an independent proof of the invariance of the pure supergravity action. 

\newpage 

\medskip
\section*{Acknowledgments.}
It was a wonderful idea of Anna Ceresole and Gianguido Dall'Agata of creating a book to celebrate 50 years of supergravity. I thank them for giving me this opportunity to contribute by writing a history of the early developments in this field. While writing this review, I realized how much had already been accomplished in the initial years. Therefore I chose to concentrate on the beginning  and  I added to the title `The first 10 years'.

I dedicate this paper to the fellow researchers of the first years. It is very sad that some of them are no longer with us to witness the 50th anniversary of Supergravity.

\newpage
\providecommand{\href}[2]{#2}\begingroup\raggedright\endgroup

\end{document}